\documentclass[twocolumn]{svjour3}          

\smartqed  
\usepackage{amsmath}
\usepackage{eucal}
\usepackage{amsfonts}
\usepackage{graphicx}
\journalname{arXiv}
\begin{document}

\title{Will claim history become a deprecated rating factor? An optimal design method on real-time road risk model
}


\author{Jiamin Yu        
}


\institute{Jiamin Yu 
		\at Shanghai Lixin University of Accounting and Finance, \\
			Shangchun Road 995, Pudong, Shanghai 201209, China  \\
              \email{iambabyface@hotmail.com}           
}

\date{Received: date / Accepted: date}

\maketitle

\begin{abstract}
With the popularity of the Telematics and Self-driving, more and more rating factors, such as mileage, route, driving behavior, etc., are introduced into actuarial models. There are quite a few doubts and disputes on the rationality and accuracy of the selection of rating variable, but it does not involve the widely accepted historical claim records. Recently, Tesla Insurance released a new generation Safety Score-based insurance, irrespective of accident history. Forward-looking experts and scholars began to discuss whether claim history will disappear in the future auto insurance ratemaking system. Therefore, this paper proposes a new risk variable elimination method as well as a real-time road risk model design framework, and concludes that claim history will be regarded as a "noise" factor and deprecated in the Pay-How-You-Drive model.
\end{abstract}

\section{Introduction}
\label{intro}
In October 2021, Tesla Insurance first released the Safety Score-based insurance, a real-time driving behavior auto insurance product. The premium of specific cover is determined by vehicle model, mileage, driving territory, and especially the {\itshape safety score}, which is calculated on five driving behavior metrics ({\itshape Forward Collision Warnings per 1,000 Miles, Hard Braking, Aggressive Turning, Unsafe Following, Forced Autopilot Disengagement}). Obviously, this new insurance is a Pay-How-You-Drive (PHYD) type of Usage-based insurance (UBI). Surprisingly, the Safety Score-based insurance not only drops traditional pricing variables (e.g. age, credit, gender, marital status, violation), but also eliminates the most important rating factor — claim (accident) history.
\paragraph{}
In practice, claim history has been the most important rate factor in auto insurance over the past 60 years and is also the cornerstone of dominant rating systems, such as No-Claim-Discount (NCD) and Bonus-Malus-System (BMS). The deprecation of claim history is not only a matter of insurance economics but also a strictly scientific and mathematical problem. For a real-time road risk rating system like the Safety Score-based insurance, we should answer the following questions.
\begin{enumerate}
	\item Is there sufficient proof for the variable elimination of claim history? \label{question:1}
	\item If the answer to (\ref{question:1}) is affirmative, whether the acturial model without claim history is the optimal risk predictive model? \label{question:2}
	\item If the answer to (\ref{question:2}) is negative, how to obtain the optimal risk predictive model by the design of observational or experimental variables? \label{question:3}
\end{enumerate}
\paragraph{}
While more and more telematics and autonomous driving data are available for rating systems, claim history is still considered to be the best predictor of future accidents in the state-of-the-art actuarial literatures. Lemaire et al. (2016) argued that annual mileage and claim history (BMS level) are two main powerful rating variables of an accurate rating system, but annual mileage should not take the place of claim history\cite{Lemaire2015THEUO}. Ma et al. (2018) and Ayuso et al. (2019) examined how to incorporate GPS trajectories and past accident data to quantify the relationship between driving habits and accidents, their findings can facilitate the development of UBI programs by extending existing experience rating systems\cite{Ma2018TheUO}\cite{Ayuso2019ImprovingAI}. Denuit et al. (2019) proposed an new multivariate credibility model that incorporates telematics data jointly with claim history to upgrade experience rating \cite{Denuit2019MultivariateCM}. Guillen et al. (2021) proposed a method to integrate near-miss telematics data into a PHYD rating system, and demostrated how to develop a basic pricing scheme based on claim history records and near‐miss observations \cite{Guilln2021NearmissTI}. Gao et al. (2022) proposed two data-driven neural network approaches to boost Poisson regression models of claim frequency with telematics data, and concluded that both classical actuarial risk factors (e.g. claim counts) and telematics data are necessary to optimize predictive models \cite{Gao2022BoostingPR}. A notable exception seems to be Yu (2021) \cite{Yu2021RiskMR}, who proposed a micro-state risk modeling method for autonomous driving scenarios, regardless of the vehicle’s claim history. Regrettably, the authors did not give any reason for variable elimination of claim history.
\paragraph{}
In this research, we attempt to answer the abovementioned three questions by applying causal inference and information theory on the basis of the existing actuarial literature. The remainder of this article is organized as follows. Section~\ref{sec:1} discusses the basic causal diagram framework of rating systems and criteria for predictive capacity of actuarial models. Section~\ref{sec:2} presents the model framwork and variable elimination methods for the optimal predictive risk model. An example of real-time road risk model is illustrated in Section~\ref{sec:3}. Section~\ref{sec:5} concludes.

\section{Methodology}
\label{sec:1}
\subsection{Actuarial criteria of variable and model selection}
\label{sec:11}
Lemaire et al. (2016) has pointed out that a strong relationship between rating variable and claims is the most important actuarial fairness criterion (accuracy). Further, we argue that {\itshape cause-effect relationship} is a stronger relationship than correlational relationship, which is the zeroth law of our study. Undoubtedly, a one-to-one functional relationship is the strongest relationship, however, this fully deterministic relationship hardly exists in actuarial models. Therefore, we practically employ {\itshape causal chain} and {\itshape causal inference} to identify and select better (more accurate) rating variables and models, so as to make these causal relationships quantifiable and computable. Different from conventional statistical variable and model selection strategies, we propose a strategy for recovering causal structure from available pricing data.
\subsection{Causal chains in auto insurance}
\label{sec:12}
In auto insurance, three types of variables are universally used for actuarial modeling, namely past historical claims, current underwriting classification variables, and future claims (as outcome variable). We denote claim history, current classification variables, and claim future as $Y_h$, $X_c$, and $Y_f$, respectively.
\paragraph{}
Note that there is a temporal precedence among $(Y_h, X_c, Y_f)$, from the past to the future. Consequently, with the theory of inferred causation (Pearl and Verma, 1995)\cite{PEARL1995ATO}, we treat claim future $Y_f$ as effects, and treat $Y_h$ and $X_c$ as causes (or indirect causes). Thus, there are four possible causal chain cases among $Y_h$, $X_c$, and $Y_f$, as below.
\begin{figure}[h]
\centering
  \includegraphics[width=0.45\textwidth]{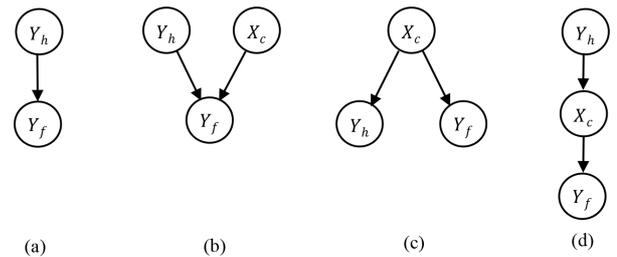}
\caption{The four possible causal chain cases: (a) A direct causal effect; (b) A common effect; (c) A common cause; (d) An indirect causal effect.}
\label{fig:1}       
\end{figure}
\\As shown in Fig.~\ref{fig:1}, possible causal chains are illustrated as directed graphs of {\itshape causal diagram} (Pearl, 1995)\cite{Pearl1995CausalDF}, where directed arrows represent causal links from cause to effect.

\paragraph{}
With these causal diagrams, we can analyze or build rating systems from a causal law perspective. This graphical method is instrumental for gaining insight into the basic philosophy behind rating systems.
\begin{enumerate}
	\item {\itshape Direct causal effect} (Fig.~\ref{fig:1}a) $\rightarrow$ {\itshape Naïve BMS}. In this simplest case, $Y_h$ as an evidence factor directly affects the occurrence of $Y_f$, which is the inherent pattern and idea of naïve BMS. 
	\item {\itshape Common effect} (Fig.~\ref{fig:1}b). In this case, both $Y_h$ and $X_c$ can influence the occurrence of $Y_f$. It seems that the addition of $X_c$’s information can enhance the naïve BMS’s prediction. In practice, however, the independence condition of $Y_h$ and $X_c$ hardly exists, because the variable selection criterion for $X_c$ is the strong correlation with claims $Y_h$. Hence, the causal diagram in Fig.~\ref{fig:1}b is rarely applied to establish rating systems.
	\item {\itshape Common cause} (Fig.~\ref{fig:1}c) $\rightarrow$ {\itshape augmented BMS}. This causal chain appears most frequently in various telematics actuarial modeling studies. In this case, the cause $X_c$ can affect both $Y_h$ and $Y_f$ outcomes. In actuarial practice, moreover, the causal link of $X_c\rightarrow Y_f$ is assumed to be identical to the causal pattern of $X_c\rightarrow Y_f$, which is the underlying assumption of numerous Poisson regression or generalized linear models (we can regard them as augmented BMS models which incorporate some priori information $X_c$, such as telematics data).
	\item {\itshape Indirect causal effect} (Fig.~\ref{fig:1}d). In this case, $Y_h$ can indirectly influence $Y_f$ via $X_c$. Compared with the common cause case in Fig.~\ref{fig:1}c, $X_c$ acts as an effect of $Y_h$, rather than a cause of $Y_h$ (or evidential chain). Few actuarial modeling studies apply this causal chain, because it requires more complex two-step inference: (a) How past evidence $Y_h$ affected $X_c$? (b) How $X_c$ will influence $Y_f$ prospectively? Obviously, the augmented BMS actuarial models may focus only one retrospective pattern of $X_c\rightarrow Y_h$, while actuarial models based on indirect causal effect shall concern another prospective pattern $X_c\rightarrow Y_f$ in addition.
\end{enumerate}

\paragraph{}
More importantly, causal diagrams provide an visually intuitional clue: {\itshape Blocking off all causal chains from $Y_h$ to $Y_f$, is a sufficient graphical proof for variable elimination of claim history $Y_h$}.

\subsection{Quantification of the predictive capacity of actuarial models}
\label{sec:13}
Let us reexamine the actuarial problem itself. The actuarial problem of auto insurance, in essence, is how to predict the unknown claim future ($Y_f$) in terms of known rating variables (represented as $X$). Clearly, optimal predictive model selection in actuarial modeling is an indispensable procedure to improve the goodness (or accuracy) of prediction. For the goal of optimal prediction, therefore, we must define and quantify the predictive capacity (goodness) of actuarial models mathematically, in other words, what is this predictive capacity of actuarial models, and how is it measured?

\subsubsection{Conventional statistical actuarial model’s criteria}
The dominating actuarial models of auto insurance are almost all based on statistical methods. Naturally, the predictive capacity of conventional actuarial models adopts statistical quantitative criteria of regression prediction.
\paragraph{}
Statistically, the prediction of claim history $Y_h$ versus rating variables $X$ is deemed as a surrogate of the prediction of claim future $Y_f$ versus $X$. Consequently, some fit statistics, such as squared error, mean absolute error, and highest posterior probability, are instinctively regarded as the measurement of predictive capacity (Barbieri and Berger, 2004)\cite{Barbieri2004OptimalPM}. Furthermore, in order to reduce possible overfitting, Akaike’s Information Criterion (AIC) and Bayesian information criterion (BIC) are commonly applied as measures of the optimal predictive model. AIC and BIC are also the most frequent model optimization criteria in actuarial modeling. In practice, for these measures, quite a few variable selection procedures (or algorithms), such as Forward Selection, Backward Elimination, or Stepwise Regression (Hocking, 1974; Beale, 1970; Mantel,1970)\cite{Hocking1976TheAA}\cite{Beale1970NoteOP}\cite{Mantel1970WhySP}, are now available for variable elimination and model optimization in the form of software packages (Lindsey and Sheather, 2010)\cite{Lindsey2010VariableSI}.
\paragraph{}
Once we choose historical claims as the dependent variable of regression model, claim history $Y_h$, directly determining claim frequency, will hardly be eliminated different than covariate $X$, which might be eliminated by variable selection procedures. In other words, claim history cannot be deprecated by conventional statistical actuarial models.

\subsubsection{The mutual information criterion}
We argue that the conventional statistical actuarial model’s criteria are flawed, because past patterns ($Y_h$ versus $X$) are not fully equivalent to future patterns ($Y_f$ versus $X$). Different from the conventional actuarial methods, we will focus more on the predictive capacity of future patterns.
\paragraph{}
By definition, both {\itshape risk} (in risk theory) and {\itshape information} (in information theory) are {\itshape uncertainties} theoretically. Naturally, we can treat the risk (or uncertainty) in rating system as information, such as risk information hidden in $X$ and $Y_f$, etc. Thus, we can also treat any rating system as a communication system from $X$ to $Y_f$, with risk information  transmitted in system channels. Under this framework, we can define the predictive capacity of such an information system as the mutual information (amount) between $X$ and $Y_f$, written as
\begin{equation}
I(X;Y_f). \notag
\end{equation}
According to the definition of mutual information (Cover and Thomas, 2006)\cite{cover2006elements}, $I(X;Y_f)$ is the reduction in prediction uncertainty (i.e. increment on prediction accuracy) of $Y_f$ after observing $X$. Consequently, $I(X;Y_f)$ can also be written as
\begin{equation}
I(X;Y_f)=H(Y_f)-H(Y_f|X) \label{eq:1}
\end{equation}
Equation (\ref{eq:1}) is the property form of mutual information, where $H(\cdot)$ represents information entropy function and the conditional entropy $H(Y_f|X)$ is the prediction uncertainty abovementioned. Note that the information measure of prediction model here, is completely different from the information measures of AIC and BIC, which are only related to the count of covariates $X$.
\paragraph{}
Through the mutual information measure, we can conclude that {\itshape the optimization goal of actuarial model is to maximize its mutual information}. Let's examine the mutual information of rating systems corresponding to the causal chains in Fig.~\ref{fig:1}.
\begin{enumerate}
	\item {\itshape Naïve BMS} (Fig.~\ref{fig:1}a). The naïve BMS predicts claim future $Y_f$ only based on claim history $Y_h$. Its mutual information can be written as
\begin{equation}
I(Y_h;Y_f)=H(Y_f)-H(Y_f|Y_h) \label{eq:2}
\end{equation}
Obviously, the prediction capacity of BMS is depends on the conditional entropy $H(Y_f|Y_h)$.
	\item {\itshape augmented BMS} (Fig.~\ref{fig:1}c). On the basis of naïve BMS, we incorporate extra priori risk information $X_c$, and then the mutual information of the augmented BMS is
\begin{equation}
I(Y_h,X_c;Y_f). \notag
\end{equation}
With the chain rule for mutual information, we have
\begin{equation}
I(Y_h,X_c;Y_f)=I(Y_h;Y_f)+I(X_c|Y_h;Y_f) \label{eq:3}
\end{equation}
Where, $I(X_c|Y_h;Y_f)$ represents the prediction capacity of $Y_f$ after observing $X_c$ under given claim history $Y_h$. Obviously $I(X_c|Y_h;Y_f)\geq 0$, then we obtain
\begin{equation}
I(Y_h,X_c;Y_f)\geq I(Y_h;Y_f) \label{eq:4}
\end{equation}
Inequality (\ref{eq:4}) shows that, at least in theory, the augmented BMS has better prediction performance than naïve BMS (Because side information $X_c$ can decrease the prediction uncertainty of BMS).
	\item {\itshape A potential rating system} (Fig.~\ref{fig:1}d). Please note that $I(Y_h,X_c;Y_f)$ has another form of the chain rule for mutual information, namely
\begin{equation}
I(Y_h,X_c;Y_f)=I(X_c;Y_f)+I(Y_h|X_c;Y_f) \label{eq:5}
\end{equation}
Where, $I(X_c;Y_f)$ is a major prediction capacity when only observing $X_c$ (such as telematics or self-driving variables, driver behaviors, etc.), and $I(Y_h|X_c;Y_f)$ represents a minor prediction capacity of $Y_f$ after observing $Y_h$ under given (or known) $X_c$. Compared with (\ref{eq:4}), $I(Y_h;Y_f)$ of BMS disappears, and $I(X_c;Y_f)$ supersedes the role of $I(Y_h;Y_f)$, which means claim history $Y_h$ might lose its dominant position in such an emerging rating system. 
\end{enumerate}
\paragraph{Significance}
In summary, (\ref{eq:5}) essentially gives out {\it the possible existence of deprecation of claim history $Y_h$ in terms of information theory}.

\section{Model design and variable elimination}
\label{sec:2}
Since Equation (\ref{eq:5}) addresses the possibility of claim history’s variable elimination, we treat the causal diagram of Fig.~\ref{fig:1}(d) as a starting point for model design. We hope to derive the independence and dependencies between risk variables and claim future from causal diagrams. Once there is independence or conditional independence between a certain risk variable and claim future, then we can conclude that there is no active causal chain path between them, that is, this risk variable is just a negligible {\itshape noise} variable versus claim future. Hence, we can use the causal diagram structure to determine whether a certain variable is a noise variable or an effective risk signal variable. Furthermore, for the optimal model design, we first define the optimal risk model criterion on the basis of our zeroth law: {\itshape the causal effects of risk variables on claim future shall be fully identifiable, that is, a fully cause-effect risk predictive model}.
\subsection{Idealized model without unobserved factors}
For the causal chain case $Y_h→X_c→Y_f$ in Fig.~\ref{fig:1}(d), we assume that there is no other common unobserved causal factors (i.e. confounders $U$) among $Y_h$, $X_c$, and $Y_f$. This idealized assumption implies that there is no back-door causal path among $Y_h$, $X_c$, and $Y_f$ (see back-door criterion in (Pearl, 1995)\cite{Pearl1995CausalDF}). Under this assumption, we can directly use the d-separation algorithm (Pearl, 1995)\cite{Pearl1995CausalDF} to obtain conditional independence between $Y_h$ and $Y_f$, namely
\begin{equation}
(Y_h\perp Y_f|X_c). \notag
\end{equation}
Here, operator $\perp$ denotes mutual independence. $(Y_h\perp Y_f|X_c)$ means that, if $X_c$ is observed, then $Y_h$ cannot influence $Y_f$ via $X_c$. If we have observed all direct causes $X_c$ of $Y_f$, we obtain
\begin{equation}
(Y_h\perp Y_f|X_c)\Longrightarrow I(Y_h|X_c;Y_f)\equiv 0	 \label{eq:6}
\end{equation}
Then, (5) becomes
\begin{multline}
I(Y_h,X_c;Y_f)=I(X_c;Y_f)+I(Y_h|X_c;Y_f)=I(X_c;Y_f)+0 \\
\equiv I(X_c;Y_f)	 \label{eq:7}
\end{multline}
As shown in (\ref{eq:7}), if we can capture all direct causes $X_c$ without confounding variable $U$, then $X_c$ will block the risk information (cause) from $Y_h$. In other words, the idealized rating system will deprecate claim history $Y_h$.

\subsection{Generalized model with unobserved factors}
More generally, due to the cost of data collection, technical feasibility or privacy, the assumption of no confounding variables is almost impossible to meet in reality. Hence, we shall consider to add unobserved factors to the causal diagram in in Fig.~\ref{fig:1}(d). There are three basic unobserved confounder structures as below.
\begin{figure}[h]
\centering
  \includegraphics[width=0.45\textwidth]{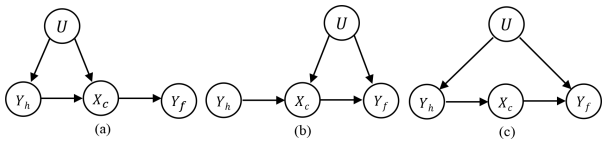}
\caption{The three unobserved confounder cases: (a) $Y_h\leftarrow U\rightarrow X_c$; (b) $X_c\leftarrow U\rightarrow Y_f$; (c) $Y_h\leftarrow U\rightarrow Y_f$.}
\label{fig:2}       
\end{figure}
\paragraph{}
For the unobserved confounder case in Fig.~\ref{fig:2}(a), we can conveniently use the {\it d-separation} algorithm to obtain conditional independences as follows:
\begin{equation}
(Y_h\perp Y_f|X_c)\quad \text{and} \quad (U\perp Y_f|X_c). \notag
\end{equation}
It means that claim history $Y_h$ and confounders $U$ can be eliminated from rating systems when $X_c$ given.
\paragraph{}
For the unobserved confounder case in Fig.~\ref{fig:2}(b), we can still use the {\it d-separation} algorithm to obtain only one conditional independence:
\begin{equation}
(Y_h\perp Y_f|X_c). \notag
\end{equation}
It means that claim history $Y_h$ can be eliminated from rating system when $X_c$ given. However, there is a {\it back-door} causal path between $X_c$ and $Y_f$, that is, $X_c$ can influence $Y_f$ via $U$ if and only if $U$ is not observed. Due to the existence of the confounding variable $U$, it is impossible to exclude the confounding bias to build the optimal rating system ($X_c$ on $Y_f$), even if claim history $Y_h$ can be eliminated. Because the prediction capacity of new rating system equals to
\begin{equation}
I(X_c;Y_f)=I(U,X_c;Y_f)-I(U|X_c;Y_f) \label{eq:8}
\end{equation}
Obviously $I(U|X_c;Y_f)>0$, $I(X_c;Y_f)$ is completely weakened by the confounding bias of $U$ (unknown). In brief, $I(X_c;Y_f)$ is only an inference fallacy of underlying rating system’s mutual information $I(U,X_c;Y_f)$. This case illustrates a point that, {\it if the causal effect of risk variables on claim future is unidentifiable, the rating model may not be the optimal risk model even if claim future can be eliminated}.
\paragraph{}
For the unobserved confounder case in Fig.~\ref{fig:2}(c), there is a back-door causal path between $X_c$ and $Y_f$ (i.e. $(Y_h\not\perp Y_f|X_c)$), which prevent us to obtain any conditional independence between $X_c$ and $Y_f$. In this case, $I(Y_h|X_c;Y_f)\ne 0$, we cannot eliminate $Y_h$. However, we can use {\it the front-door criterion} (Pearl, 1995)\cite{Pearl1995CausalDF} to obtain $I(Y_h,X_c;Y_f)$ with ignoring the existence of unknown $U$. Note that the elimination of unobserved confounders $U$ here is achieved by the {\it do-calculus} (or {\it intervention-calculus}, see (Pearl, 1995)\cite{Pearl1995CausalDF} ) operations: $do(Y_h)$ and $do(X_c)$.

\subsection{Optimal model with unobserved factors}
Although the cases in Fig.~\ref{fig:2}(b) and \ref{fig:2}(c) are not the optimal model, they provide clues for building the optimal model. We can combine them to eliminate not only claim history $Y_h$, but also the confounding bias of $U$, so as to make the causal effect of $X_c$ on $Y_h$ fully identifiable . The new optimal model has the following causal diagram.
\begin{figure}[h]
\centering
  \includegraphics[width=0.3\textwidth]{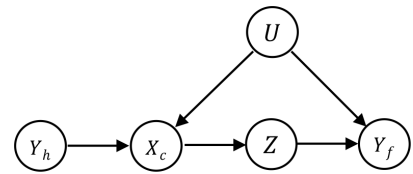}
\caption{The causal diagram of optimal model with unobserved $U$.}
\label{fig:3}       
\end{figure}
\\Where, $Z$ is a new deliberately designed, additional observation variables with direct causal link to $X_c$ and $Y_f$.
\paragraph{}
As shown in Fig.~\ref{fig:3}, the optimal model integrates two merits: One is that the variable elimination of $Y_h$ can be realized by the {\it d-separation} algorithm, and the other is that the interference of unobserved confounding variable $U$ can be avoided by the {\it front-door criterion} (i.e. the causal effect of $X_c$ on $Y_f$ is identifiable and computable).
\paragraph{}
In the next section, we will demonstrate how the framework of Fig.~\ref{fig:3} above can be used to derive a real-time road risk model, with stronger causality. 

\section{Real-time road risk model: an example}
\label{sec:3}
In this section, we will illustrate the framework abovementioned through a real-time road risk model example on self-driving (Yu, 2021)\cite{Yu2021RiskMR}. Yu (2021) considers the accident occurrence mechanism as an ergodic process of perilous states, which is from the {\it safe-state}, via the {\it contact-state}, the {\it conflict-state}, to the {\it crash-state}. By means of a safety indicator, $TTA$ (time-to-accident), these driving states can be further discretized into a series of driving state events, denoted as $S_0,S_1,\cdots,S_{D+1}$ ($S_0$ and $S_{D+1}$ represent absolutely safe and accident events respectively). With such an accident mechanism, we can construct a corresponding causal diagram of driving event flow as below.
\begin{figure}[h]
\centering
  \includegraphics[width=0.48\textwidth]{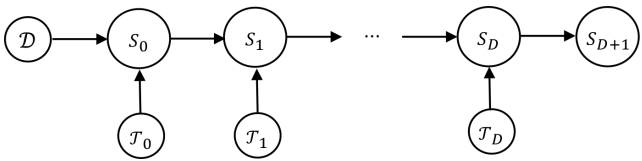}
\caption{The causal diagram of real-time driving events.}
\label{fig:4}       
\end{figure}
\\Where, $\mathcal{D}$ represents a driving decision set of current self-driving AI or human driver, and $\mathcal{T}_i$ represents the instant traffic condition of driving event $S_i$, while solid arrows represent causal links among these quantities. Obviously, $\mathcal{D}$ belongs to self-driving vehicle or human factors, and $\mathcal{T}_i$ belongs to road environment factors.
\paragraph{}
Assume that there is no other unobserved variable in Fig.~\ref{fig:4}, we can apply {\it d-separation} to obtain the recursive decomposition of total accident probability given $\mathcal{D}$:
\begin{equation}
P(S_0,S_1,\cdots,S_{D+1}|\mathcal{D})=\prod_{i=0}^D{P(S_{i+1}|S_i,\mathcal{D})} \label{eq:9}
\end{equation}
\\Because $S_i$ is derived and refined from current traffic observed data $\mathcal{T}_i$ (e.g. traffic velocity, density, obstacle detection, etc.), $\mathcal{T}_i$ will be blocked any causal path to $S_{i+1}$ when $S_i$ is observed or measured. In other words, $S_i$ can d-separate $\mathcal{T}_i$ from $S_{i+1}$, written $(\mathcal{T}_i\perp S_{i+1}|S_i)$. Thus it can be seen that the conditional independence $(\mathcal{T}_i\perp S_{i+1}|S_i)$ is the underlying causality philosophy of Yu (2021).

\subsection{Basic framework}
Combining the causal diagrams in Fig.~\ref{fig:3} and \ref{fig:4}, we can derive a new causal diagram framework for the optimal real-time model. First, the final accident state $S_{D+1}$ is equivalent to claim future $Y_f$, that is, a crash is assumed to definitely trigger a claim only for the sake of model’s simplicity. Second, for a self-driving vehicle, its driving decision set $\mathcal{D}$ can be measured and calibrated in advance. Obviously, we can regard $\mathcal{D}$ as current classification variables $X_c$. Third, the measurable state set $\mathcal{S}={S_0,\cdots,S_D}$ in Fig.~\ref{fig:4}, is the equivalent of observed variable set $Z$. With grafting Fig.~\ref{fig:4} into Fig.~\ref{fig:3}, we have an upgraded causal diagram framework as below.
\begin{figure}[h]
\centering
  \includegraphics[width=0.35\textwidth]{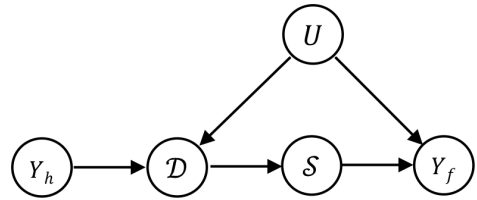}
\caption{The causal diagram of optimal model for real-time driving events.}
\label{fig:5}       
\end{figure}
\paragraph{}
It is worth noting that the unobserved confounder $U$ widely exists in real road driving scenarios with a high probability. For instance, encountering road rage drivers or sudden traffic offenders, can directly influence and change both driving decisions $\mathcal{D}$ and accident occurrence $Y_f$.

\subsection{Controlling confounding bias from $U$}
Fortunately, the measurable state set $\mathcal{S}$ is composed of experimental design covariates $\{S_i\}$ (Yu, 2021). This means there is an unique path from $\mathcal{D}$ to $Y_f$, via $S_0,S_1,\cdots,S_D$, which can be easily proved by the time continuity of $TTA$ (i.e. state $S_{i-1}$ cannot bypass state $S_i$ to enter state $S_{i+1}$). Thus, in Fig.~\ref{fig:5}, the applicable conditions of the front-door criterion theorem are satisfied: (a) $\mathcal{S}$ intercepts all directed paths from $\mathcal{D}$ to $Y_f$, (b) there is no back-door path between $\mathcal{D}$ and $\mathcal{S}$, and (c) every back-door path between $\mathcal{S}$ and $Y_f$ is blocked by $\mathcal{D}$. Using the front-door criterion theorem, we can eliminate the unobserved confounder $U$ to obtain
\begin{equation}
P(Y_f|do(\mathcal{D}),Y_h)=\sum_\mathcal{S}{P(\mathcal{S}|\mathcal{D})\sum_{\mathcal{D}'}{P(Y_f|\mathcal{D}',\mathcal{S})P(\mathcal{D}'|Y_h)} } \label{eq:10}
\end{equation}
\\Where, $P(Y_f|do(\mathcal{D}),Y_h)$ represents the causal effect of driving decisions $\mathcal{D}$ on claim future $Y_f$ given claim history $Y_h$.
\paragraph{}
Front-door formula (\ref{eq:10}) means that we can identify and calculate the causal-effect relationships between driving decisions and claim future by measuring designed microscopic risk indicators (covariates), such as $\{S_i\}$.

\subsection{Variable elimination of historical claims $Y_h$}
It is not difficult to find that, from historical accidents to activation of driving strategies, and even to future accidents, it is inevitable to experience one journey start event. Such a journey start event can be regarded as a switch variable, which determines whether an accident occurs afterwards (i.e. an accident will happen if and only if one hits the road). We denote this switch event variable as $J_o$, one element of current classification variables $X_c$. Then, the optimal model in Fig.~\ref{fig:5} turns into
\begin{figure}[h]
\centering
  \includegraphics[width=0.45\textwidth]{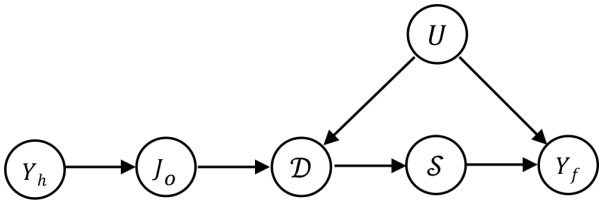}
\caption{The causal diagram of optimal model with journey start events.}
\label{fig:6}       
\end{figure}
\paragraph{}
When $J_o$ is measurable, we can use inference Rule 1 ({\it insertion/deletion of observations}) (see Theorem 3 in Pearl (1995)) to ‘wipe out’ claim history $Y_h$. Because the conditional independence $(Y_f\perp Y_h |J_o,\mathcal{D})$ holds, we have following
\begin{multline}
P(Y_f|do(J_o),do(\mathcal{D}),Y_h)=P(Y_f|do(J_o),do(\mathcal{D}))\\
=P(Y_f|do(\{J_o,\mathcal{D}\})) \label{eq:11}
\end{multline}
\\As shown in Fig.~\ref{fig:6}, (\ref{eq:11}) reaffirms that $Y_f$ and $Y_h$ can be {\it d-separated} by an intentionally designed event variable $J_o$. Once $J_o$ is observed, $Y_f$ and $Y_h$ are mutually independent, thereby $Y_h$ can be eliminated from rating systems as noisy signal observations.

\subsection{A framework of the optimal PHYD model}
After controlling confounding bias from $U$ and variable elimination of $Y_h$, we can obtain a new causal diagram of the optimal model as following.
\begin{figure}[h]
\centering
  \includegraphics[width=0.35\textwidth]{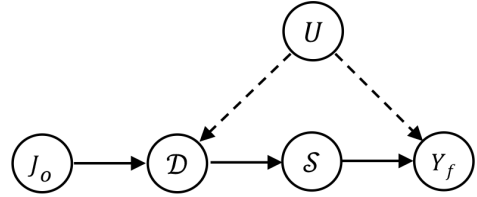}
\caption{The causal diagram of optimal PHYD model with driving behaviors $\{J_o,\mathcal{D}\}$.}
\label{fig:7}       
\end{figure}
\\Where dashed arrows represent controlled (negligible) confounding links. Intuitively, this new structure of Fig.~\ref{fig:7} permits us to identify and quantify the causal effects of $\{J_o,\mathcal{D}\}$ on $Y_f$. The set $\{J_o,\mathcal{D}\}$ just exactly indicates the measurements of driving activity (exposure) and driving decision behaviors. Thereby, the framework in Fig.~\ref{fig:7} is precisely a PHYD model with driving behaviors.
\paragraph{}
Combining (\ref{eq:10}) and (\ref{eq:11}), we can obtain the causal effect $P(Y_f|do(\{J_o,\mathcal{D}\}))$ the of optimal PHYD model in Fig.~\ref{fig:7}.
\begin{multline}
P(Y_f|do(\{J_o,\mathcal{D}\}))\\
=\sum_\mathcal{S}{P(\mathcal{S}|\mathcal{D})\sum_{\mathcal{D}'}{P(Y_f|\mathcal{D}',\mathcal{S})P(\mathcal{D}'|do(J_o))} } \label{eq:12}
\end{multline}
\\Here, for the sake of model simplification and demonstration, we assume that causal effect $P(\mathcal{D}'|do(J_o))$ of $J_o$ on $\mathcal{D}$ is equal to $P(\mathcal{D}'|J_o)$. Then, we have
\begin{equation}
P(Y_f|do(\{J_o,\mathcal{D}\}))=\sum_\mathcal{S}{P(\mathcal{S}|\mathcal{D})\sum_{\mathcal{D}'}{P(Y_f|\mathcal{D}',\mathcal{S})P(\mathcal{D}'|J_o)} } \label{eq:13}
\end{equation}
\\Formula (\ref{eq:13}) provides a calculation method on the indirect causal effect of driving behaviors $\{J_o,\mathcal{D}\}$ on accidents $Y_f$, by {\it surrogate traffic engineering indicators} $\mathcal{S}$.

\section{Conclusions}
In this study, we apply information theory and causal inference to introduce three actuarial modeling ideas: (a) Causal diagram can be used to discover the paradigm of causality hidden in rating systems; (b) The optimization of actuarial predictive model is to maximize the mutual information of the rating system; (c) The algorithms and criteria of causal diagram are powerful tools to derive the optimal risk predictive model. 
\paragraph{}
With these methodologies and ideas, we start to present a model design framework of the optimal predictive model as well as a variable elimination method. To illustrate the framework and method, we further to propose an optimal model design instance of real-time road risk, with variable elimination of claim history $Y_h$. Meanwhile, we obtain our solutions to three fundamental problems proposed in the Introduction section: (a) The conditional independence from d-separation is a sufficient proof for variable elimination of claim history $Y_h$; (b) The identifiability of the causal effect of risk factors $X_c$ on claim future $Y_f$, is the optimality criterion of actuarial modeling; (c) The optimal risk predictive model shall incorporate experimental design variables beyond observational variables, to ensure the causal effect of $X_c$ on $Y_f$ is identifiable.
\paragraph{}
The model design framework developed in this paper will directly facilitate actuarial modeling of UBI or PHYD, and even provide decision-making tools for auto accident prevention and mitigation. The work also provides a new research and development interface for traffic engineering and vehicle engineering specialists to participate in risk modeling and management.


%

\bibliographystyle{siam}   		
\bibliography{bibrefs}   			

%
%

\end{document}